\documentclass[pra,showpacs,twocolumn,showkeys]{revtex4-1}
\usepackage{graphics}
\usepackage{amsmath}
\usepackage{color}
\usepackage{braket}
\usepackage{dsfont}
\newcommand{\enquote}[1]{``#1''}
\begin{document}
\newcommand{\bibs}{d:/Dropbox/Dad/Mark/References/BibFile}
\newcommand{\bstfile}{osa}
\title{Quantum no-cloning theorem and entanglement}
\author{Mark G. Kuzyk}
\affiliation{Department of Physics and Astronomy, Washington State University, Pullman, Washington  99164-2814}

\maketitle

The article by Ortigoso\cite{ortig18.01} describes the history of the quantum no-cloning theorem,\cite{Wooters82.01,dieks82.01} arguing that Jim Park discovered it 12 years prior to 1982.\cite{park70.01}  Given that AJP is dedicated to pedagogy, I am writing to recast the state vectors given by Ortigoso in a form that reflects the required statistics of indistinguishable particles, which might be less confusing to students.  This more precise form also applies to entanglement, which I believe is often described in a way that implies that identical particles may sometimes be excused from spin statistics.  Here I describe both.

Cloning means making an exact copy, where two identical systems result; the original and the copy.  Consider a single particle in state $\Ket{\phi}$ that is cloned, leaving the two-particle state $\Ket{\phi}\Ket{\phi}$.  This state is nonsensical for a fermion, which forbids two particles from occupying the same state.

The resolution is to start with a particle {\em here} whose state we wish to duplicate {\em there}.  Rather than expressing the initial state of the two identical particles as $\Ket{\phi} \Ket{0}$, where the state $\Ket{\phi}$ is to be copied according to the operation $T \Ket{\phi} \Ket{0} \rightarrow \Ket{\phi} \Ket{\phi}$, the initial two-particle state $\Ket{\phi_h,0_t}$ with exchange symmetry is given by
\begin{align}\label{eq:initial}
\Ket{\phi_h,0_t} \equiv \frac { \Ket{\phi_h} \Ket{0_t} \pm \Ket{0_t} \Ket{\phi_h} } { \sqrt{2}},
\end{align}
where the subscripts $h$ and $t$ refer to {\em here} and {\em there} and the $\pm$ signs designate bosons/fermions.  The first ket represents Particle \#1 and the second ket Particle \#2.  Note that we use the convention that $\Ket{\phi_h} \Ket{0_t}$ represents a direct product while $\Ket{\phi_h,0_t}$ is the symmetrized form as given by Eq.~(\ref{eq:initial}).  The two occupied states can be identical for bosons, such as $\Ket{\phi,\phi} = \Ket{\phi} \Ket{\phi}$,  but will generally be different.

Eq.~(\ref{eq:initial}) implicitly assumes that states here are orthogonal to states there, or $\Braket{\phi_h | \phi_t} = 0$.  If here and there represent two spatial regions separated by an infinite barrier, then the two locations span a space that is orthogonal to the internal states, so $\Ket{\phi_h} $ can be separated into a direct product of the form $\Ket{\phi_h} = \Ket{\phi} \Ket{h}$.  Then,  Eq.~(\ref{eq:initial}) can be expressed as
\begin{align}\label{eq:initialTwoSpaces}
\Ket{\phi_h,0_t} &= \frac { \Big(\Ket{\phi} \Ket{h} \Big)_1  \Big(\Ket{0} \Ket{t} \Big)_2 \pm  \Big(\Ket{0} \Ket{t} \Big)_1 \Big(\Ket{\phi} \Ket{h} \Big)_2 } { \sqrt{2}} \nonumber \\
 &= \frac { \Ket{\phi}_1 \Ket{0}_2 \Ket{h}_1 \Ket{t}_2 \pm  \Ket{0}_1 \Ket{\phi}_2 \Ket{t}_1 \Ket{h}_2 } { \sqrt{2}},
\end{align}
where the subscripts $1$ and $2$ remind us that the direct product within parentheses reefers to Particle \#1 and \#2, respectively; and where we have regrouped the terms on the last line.

When cloning yields two particles in the same ``internal" state but at two different locations, the definition of the cloning operation must be
\begin{align}\label{eq:final}
T \frac { \Ket{\phi_h} \Ket{0_t} \pm \Ket{0_t} \Ket{\phi_h} } {\sqrt{2}} \rightarrow \frac{ \Ket{\phi_h} \Ket{\phi_t} \pm \Ket{\phi_t} \Ket{\phi_h}} {\sqrt{2}} .
\end{align}
Eqs.~(\ref{eq:initial}) and (\ref{eq:final}) define the action of the cloning operation to be
\begin{align}\label{eq:clone}
T \Ket{\phi_h,0_t} =\Ket{\phi_h,\phi_t} ,
\end{align}
which does not imply that $T \Ket{\phi_h} \Ket{0_t} \rightarrow \Ket{\phi_h} \Ket{\phi_t} $.  If this were so,
\begin{align}\label{eq:FinalWrong}
T \frac { \Ket{\phi_h} \Ket{0_t} \pm \Ket{0_t} \Ket{\phi_h} } { \sqrt{2} } \rightarrow \frac { \Ket{\phi_h} \Ket{\phi_t} \pm \Ket{0_t} \Ket{0_h} } { \sqrt{2} } ,
\end{align}
which clearly is not a clone.  As such, the cloning operator cannot act on just a single direct product of state vectors but must operate on the symmetrized state vector.

The cloned state in Eq.~(\ref{eq:final}), written in the form given by Eq.~(\ref{eq:initialTwoSpaces}), separates into the product
\begin{align}\label{eq:FinalTwoSpaces}
\Ket{\phi_h,\phi_t} = \Ket{\phi}_1 \Ket{\phi}_2 \frac { \Ket{h}_1 \Ket{t}_2 \pm \Ket{t}_1 \Ket{h}_2 } { \sqrt{2}}.
\end{align}
Eq.~(\ref{eq:FinalTwoSpaces}) illustrates how the form $\Ket{\phi} \Ket{\phi} $ generally used by researchers refers to the internal part of the state vector and the exchange symmetry is contained in the second term, which is not explicitly stated but implicitly assumed.  However, the initial state vector prior to cloning given by Eq.~(\ref{eq:initialTwoSpaces}) cannot be separated into an internal and external part, so $\Ket{\phi} \Ket{0}$ is nonsensical.  As such we proceed with the initial state given by Eq.~(\ref{eq:initial}) and final state given by Eq.~(\ref{eq:clone}), and show that the proof of the no-cloning theorem parallels the standard one, as follows.

Transmitting a state $a \Ket{\psi_h} + b \Ket{\phi_h}$ {\em here} to {\em there}, which is initially in state $\Ket{0_t}$, yields
\begin{align}\label{eq:FinalSuper}
T \Ket{a\psi_h + b\phi_h, 0_t} \rightarrow  \Ket{a\psi_h + b\phi_h, a\psi_t + b\phi_t} .
\end{align}
Expanding the righthand sides of Eq.~(\ref{eq:FinalSuper}) using the definition given by Eq.~(\ref{eq:initial}) gives
\begin{align}\label{eq:FinalSuper2}
& \Ket{a\psi_h + b\phi_h, a\psi_t + b\phi_t} \nonumber \\
& = a^2 \Ket{\psi_h, \psi_t} +  ab \Big( \Ket{\psi_h, \phi_t} + \Ket{\phi_h, \psi_t} \Big)+  b^2 \Ket{\phi_h, \phi_t}.
\end{align}

The linear operator $T$ acting on the left-hand side of Eq.~(\ref{eq:FinalSuper}( according to Eq.~(\ref{eq:final}) yields
\begin{align}\label{eq:FinalSuper3}
T & \Big( a \Ket{\psi_h, 0_t} + b \Ket{\phi_h, 0_t} \Big) =   a T \Ket{\psi_h, 0_t} + b T \Ket{\phi_h, 0_t}\nonumber \\
& \rightarrow a \Ket{\psi_h, \psi_t} +  b \Ket{\phi_h, \phi_t}.
\end{align}
Eqs.~(\ref{eq:FinalSuper2}) and (\ref{eq:FinalSuper3}) are of the same form as Eqs.~(2) and (1) in Ortigoso's paper; if the righthand sides are to be equal, either $a$ or $b$ must vanish, so cloning is disallowed for a superposition of states.

Entanglement as introduced in textbooks and used extensively in the literature also appears to be excused from spin statistics.  For example, in a recent issue of AJP, Brody and Selton express the state of two photons as\cite{brody18.01}
\begin{align}\label{eq:EntangPhotons}
\Ket{\psi} = \frac {\Ket{H}_1 \Ket{H}_2 + \Ket{V}_1 \Ket{V}_2 } {\sqrt 2} ,
\end{align}
where $\Ket{H}$ and $\Ket{V}$ are horizontally- and vertically-polarized photons produced in the nonlinear-optical process of down conversion and the subscripts represent their momneta.  Brody and Selton reference a paper by Dehlinger and Mitchell\cite{dehlin01.01} for the meaning of Eq.~(\ref{eq:EntangPhotons}), which states, ``The photons are heading in different directions, and thus can be treated as distinguishable particles."  Formally, the calculations give the right answers, but the interpretation is confusing; the spin-statistics theorem requires that the bosonic state vector remain unchanged upon particle interchange,\cite{duck98.01} which is not the case here.

The resolution follows along the same line as in the no-cloning case.  Eq.~(\ref{eq:EntangPhotons}) can be forced to have the correct exchange symmetry by expressing it as
\begin{align}\label{eq:EntangPhotonsCorrect}
\Ket{\psi} = \frac {\Ket{H_1 , H_2} + \Ket{V_1 , V_2 }} {\sqrt 2} ,
\end{align}
where each of the two terms in the numerator of Eq.~(\ref{eq:EntangPhotonsCorrect}) are of the form defined by Eq.~(\ref{eq:initial}) and thus are fully entangled and have the correct symmetry.  Here, the subscripts 1 and 2 refer to the momenta of the particles.  If a measurement is made that finds a horizontally polarized photon with momentum `1', the state vector will collapse into the correctly symmetrized state $\Ket{H_1 , H_2}$.  A subsequent measurement of the polarization of the photon with momentum `1' will always find a horizontally-polarized photon there and a measurement of the photon with momentum `2' will always find a horizontally polarized photon.

The difference between $\Ket{\phi,\psi}$ and $\Ket{\phi} \Ket{\psi}$ can be visualized and reconciled by comparing the probability densities represented by each case for Particle \#1 in a two-particle system where they are made ``indistinguishable" by virtue of being localized and spatially separated.  For the symmetrized state $\Ket{\phi,\psi}$ the probability density for Particle \#1 is
\begin{align}\label{eq:prob1}
\rho(x_1) &= \left| \Braket{x_1 | \phi,\psi} \right|^2 = \left|  \frac {\phi(x_1) \Ket{\psi} \pm \psi(x_1) \Ket{\phi}} {\sqrt{2}} \right|^2 \nonumber \\
& = \frac {\left| \phi(x_1) \right|^2 + \left| \psi(x_1) \right|^2} {2} ,
\end{align}
where we have used the fact that states $\Ket{\phi}$ and $\Ket{\psi}$ are orthogonal and $\Bra{x_1}$ acts only on the first ket of a direct product.  The probability density of finding Particle \#2 at position $x_2$ is the same, that is, $\rho(x_2) = \rho(x_1)\big|_{x_1 = x_2}$, or
\begin{align}\label{eq:prob21}
\rho(x_2) = \frac {\left| \phi(x_2) \right|^2 + \left| \psi(x_2) \right|^2} {2} ,
\end{align}
reflecting the fact that a property of each of two indistinguishable particles should be the same.

Eqs.~(\ref{eq:prob1}) and (\ref{eq:prob21}) illustrate the impossibility of knowing which particle is in which state; all we can know is that one particle is in each of the two states and that exchanging the two particles does not change the probability density.

The probability density for Particle \#1 in the state $\Ket{\phi} \Ket{\psi}$ is
\begin{align}\label{eq:prob2}
\rho(x_1) &= \left| \left( \Braket{x_1 | \phi} \right) \Ket{\psi} \right|^2 = \left| \phi(x_1) \right|^2
\end{align}
and for Particle \#2 is
\begin{align}\label{eq:prob2b}
\rho(x_2) &= \left| \left(  \Ket{\phi} \Braket{x_2 | \psi} \right) \right|^2 = \left| \psi(x_2) \right|^2 .
\end{align}
In this case, the probability densities for each of the particles are different because $\Ket{\phi} \Ket{\psi}$ is not entangled, implying that the particles are distinguishable.

Eqs.~(\ref{eq:prob1}) and (\ref{eq:prob2}) are different, but they will be equivalent (as will Eqs.~(\ref{eq:prob21}) and (\ref{eq:prob2b}) when each particle is trapped in its own infinite well.  Then, each particle lives in its own Hilbert space because the wave function $\psi(x_1)$ vanishes at all locations where $\phi(x_1)$ is nonzero, and vice versa.  Then, for example, we can associate Particle \#1 with the probability density $\phi(x_1)$ in the well where it is localized and Particle \#2 with the probability density $\psi(x_2)$ where it is localized.

The probability density of Particle \#1 is found by measuring the particle's position in the well it occupies.  Since the wave functions of the particles do not overlap, restricting the measurement to one well isolates the desired probability density.  For Particle \#1, Eq.~(\ref{eq:prob1}) with $\psi(x_1) = 0 $ gives a density $\rho(x_1) = \left| \phi(x_1)\right|^2$ after re-normalizing the wavefucntion within the single well.  The density obtained is the same as given by Eq.~(\ref{eq:prob2b}).  In this case, the language of distinguishable particles is simpler, and that is why it is so often used.

For overlapping functions, Eqs.~(\ref{eq:prob1}) and (\ref{eq:prob21}) describe the probability densities and Eqs.~(\ref{eq:prob2}) and (\ref{eq:prob2b}) no longer hold.  Even in the case of two infinite wells, an individual particle's wave function can span both wells, thus requiring the language of entanglement.  Dehlinger and Mitchell's example of two counter-prorogating photons, corresponds to two non-overlapping delta functions in momentum space.  Consequently, a distinct momentum is associated with each particle in analogy to a particular well being occupied by a particular particle.

To conclude, the symmetrized state vector $\Ket{\phi,\psi}$ applies more broadly than $\Ket{\phi} \Ket{\psi}$, which holds only when the wave functions do not overlap.  To some, it may be more pleasing that $\Ket{\phi,\psi}$ embodies the correct exchange symmetry, even when particles live in disjoint spaces.  To others, convenience is paramount.  Regardless of preference, to minimize confusion, the teacher should stress the distinction between $\Ket{\phi,\psi}$ and $\Ket{\phi} \Ket{\psi}$ when discussing spin statistics and entanglement.

\end{document}